\documentclass[5p,times,procedia]{elsarticle}
\usepackage{ecrc}

\volume{00}

\firstpage{1}

\journalname{Procedia CIRP}

\runauth{Author name}

\jid{trpro}

\usepackage{amssymb}

\usepackage[figuresright]{rotating}

\usepackage[bookmarks=false]{hyperref}
    \hypersetup{colorlinks,
      linkcolor=blue,
      citecolor=blue,
      urlcolor=blue}
      
\usepackage{enumitem}

\usepackage{tikz}
\usetikzlibrary{decorations.pathmorphing}

\usepackage{adjustbox}

\usepackage[nolist]{acronym}

\begin{document}

\begin{acronym}
  \acro{dt}[DT]{\textit{Digital Twin}}
  \acroplural{dt}[DTs]{\textit{Digital Twins}}

  \acro{rq}[RQ]{\textit{Research Question}}
  \acroplural{rq}[RQs]{\textit{Research Questions}}
  
  \acro{rh}[RH]{\textit{Research Hypothesis}}
  \acroplural{rh}[RHs]{\textit{Research Hypotheses}}

  \acro{fmi}[FMI]{\textit{Functional Mockup Interface}}

  \acro{fmu}[FMU]{\textit{Functional Mockup Unit}}

  \acro{oem}[OEM]{\textit{Original Equipment Manufacturer}}
  \acroplural{oem}[OEMs]{\textit{Original Equipment Manufacturers}}

  \acro{iot}[IoT]{\textit{Internet of Things}}

  \acro{ai}[AI]{\textit{Artificial Intelligence}}

  \acro{dm}[DM]{\textit{Digital Model}}
  \acroplural{dm}[DMs]{\textit{Digital Models}}
  
\end{acronym}

\begin{frontmatter}

\dochead{19th CIRP Conference on Intelligent Computation in Manufacturing Engineering (CIRP ICME ‘25)}%

\title{An Expert Survey on Models and Digital Twins}

\author[a]{Jonathan Reif} 
\author[b]{Daniel Dittler}
\author[a]{Milapji Singh Gill}
\author[c]{Tamás Farkas}
\author[b]{Valentin Stegmaier}
\author[a]{Felix Gehlhoff}
\author[c]{Tobias Kleinert}
\author[b]{Michael Weyrich}

\address[a]{Institute of Automation Technology, Helmut Schmidt University / University of the Federal Armed Forces Hamburg, Holstenhofweg 85, 22043 Hamburg, Germany}
\address[b]{Institute of Industrial Automation and Software Engineering, University of Stuttgart, Pfaffenwaldring 47, 70569 Stuttgart, Germany}
\address[c]{Chair of Information and Automation Systems, RWTH Aachen University, Turmstraße 46, 52064 Aachen, Germany}

\aucores{* Corresponding author. Tel.: +49-040-6541-3372 ; fax: +49-040-6541-2004. {\it E-mail address:} jonathan.reif@hsu-hh.de}

\begin{abstract}

Digital Twins (DTs) are becoming increasingly vital for future industrial applications, enhancing monitoring, control, and optimization of physical assets. This enhancement is made possible by integrating various Digital Models (DMs) within DTs, which must interoperate to represent different system aspects and fulfill diverse application purposes. However, industry perspectives on the challenges and research needs for integrating these models are rarely obtained. Thus, this study conducts an expert survey across multiple application domains to identify and analyze the challenges in utilizing diverse DMs within DTs. 
The results reveal missing standardized interfaces, high manual adaptation effort, and limited support for model reuse across lifecycle phases, highlighting future research needs in automated model composition and semantics-based interoperability.
\end{abstract}


\end{frontmatter}

\section{Introduction}
\label{sec:introduction}
The concept \ac{dt} has garnered significant attention in recent years due to its potential to bridge the gap between physical systems and their digital representations, enhancing real-time monitoring, control, and optimization across various industries \cite{Kritzinger.2018, Reinpold.2024}. 
\acp{dt}, often composed of complex \acp{dm}, represent virtual counterparts of physical assets, systems, or processes, facilitating a deeper understanding of behavior across the entire lifecycle of the system \cite{Stark.2019}.

While numerous systematic literature reviews have explored the concept of \acp{dt} \cite{Kritzinger.2018, Dalibor.2022}, research still lacks a systematic capture of industry experts' perspectives on the practical application, interoperability, and composition of \acp{dm}.
This study seeks to address this gap by conducting a structured survey with industry practitioners and academic experts to systematically analyze the types and roles of \acp{dm} in industrial practice. 

To clarify the role and utility of \acp{dm} across the industrial lifecycle, this study addresses two primary \acp{rq}:

\vspace{0.2cm}
\begin{itemize}[leftmargin=*, nosep]
    \item[] \textbf{RQ 1:} \textit{What types of \acp{dm} are used within DTs across different lifecycle phases?}
    \item[] \textbf{RQ 2:} \textit{For which industrial use cases does the interoperability and composition of \acp{dm} offer significant potential?}
\end{itemize}
\vspace{0.2cm}

By combining industry and academic perspectives, this study captures current practices, key challenges, and future needs for \acp{dm} within \acp{dt} across diverse application domains. It aims to identify critical gaps for future research and assess the practical potential of model interoperability and composition to enhance lifecycle management and digital services. The results also reveal industry-specific opportunities to leverage \acp{dm} more effectively.

The remainder of this paper is structured as follows. Sec. ~\ref{sec:Related Work} provides an overview and critical analysis of related literature to contextualize the study. The applied methodology, including the survey design and participant selection, is detailed in Sec. ~\ref{sec:methodology}. Subsequently, Sec. ~\ref{sec:results} presents and analyzes the empirical findings of the expert survey, which are further discussed with regard to their implications for future \ac{dt} developments in industrial contexts. The paper concludes in Sec. ~\ref{sec:conclusion} with a summary of the key contributions and recommendations for future research.
\section{Related Work} 
\label{sec:Related Work}

The following Sec. \ref{sec:Related Work} reviews existing \ac{dt} surveys, focusing on their objectives and distinguishing them from the scope and contribution of the present study.

\citet{Kritzinger.2018} conducted a comprehensive literature review to clarify the concept of \acp{dt} in manufacturing. 
They introduced a classification framework that differentiates between \acp{dm}, Digital Shadows, and \acp{dt}, based on the direction and automation of data flow between physical and digital entities. 
Their analysis revealed that many systems labeled as \acp{dt} lacked true bidirectional, automated data exchange, often functioning as \acp{dm} or Digital Shadows.

\citet{Fuller.2020} provide a comprehensive assessment of the \ac{dt} concept, focusing on its enabling technologies, challenges, and open research questions. 
They define a \ac{dt} as the seamless integration of data between a physical and virtual system, facilitating bidirectional data flow. 
The study explores key enabling technologies such as the \ac{iot}, \ac{ai}, and data analytics, highlighting their roles in the development and implementation of \acp{dt} across various sectors, including manufacturing, healthcare, and smart cities. 
Additionally, the authors identify challenges related to standardization, data security, and the need for robust IT infrastructures to support \ac{dt} deployments.
 
\citet{Mihai.2022} provide an extensive survey on \acp{dt}, focusing on their foundational technologies, existing challenges, emerging trends, and future directions. 
The authors delve into the core technologies that facilitate \acp{dt}, such as the \ac{iot}, \ac{ai}, and data analytics, and discuss their integration across various industries. 
They also address critical challenges, including the need for standardization, ensuring data security, and managing the complexity of integrating diverse systems. 

\citet{Hildebrandt.2024} conducted a systematic literature review focusing on data integration within \acp{dt} in industrial automation. 
Analyzing 141 out of 1,107 unique publications, they identified key challenges such as the lack of standardized data models, interoperability issues among heterogeneous systems, and the complexities of real-time data processing. 
The study emphasizes the necessity for unified data integration frameworks to enhance the efficiency and scalability of \ac{dt} applications in industrial settings. 

\citet{Reinpold.2024} present a systematic comparison of Software Agents and \acp{dt} within industrial production environments. 
Their study examines the distinct roles, overlapping functionalities, and potential synergies between these two paradigms. 
The authors highlight that Software Agents are typically employed for collaborative planning and execution of production processes, leveraging their autonomy and decision-making capabilities. 
In contrast, \acp{dt} are primarily utilized for monitoring and information processing of production resources, providing high-fidelity digital representations of physical assets. 
Despite these characteristic roles, the study finds that a clear-cut distinction between the two paradigms is challenging, suggesting that integrating Software Agents and \acp{dt} could enhance system intelligence, autonomy, and sociability.

\citet{Dalibor.2022} conducted a cross-domain systematic mapping study to investigate the multifaceted nature of \acp{dt} from a software engineering perspective. 
Analyzing 356 publications across various domains, they examined the types of research contributions, expected properties, realization methods, operational aspects, and evaluation techniques associated with \acp{dt}. 
Their study culminated in the development of a novel feature model aimed at guiding future software engineering efforts in the development, deployment, and management of \acp{dt}. 

Although these studies provide significant contributions to understanding \acp{dt}, none of them systematically incorporate the perspectives of industry practitioners. 
Some works, such as those by \citet{Hildebrandt.2024} and \citet{Mihai.2022}, address interoperability challenges, but their focus is primarily on data integration rather than the composition of \acp{dm} within \acp{dt}. 
Other studies, such as those by \citet{Reinpold.2024} and \citet{Dalibor.2022}, investigate the relationship between \acp{dt} and complementary technologies, yet they remain centered on conceptual and technical frameworks rather than practical implementation challenges.

Our study aims to bridge this gap by directly engaging with experts from industry and academia to systematically assess the current state of \ac{dm} usage. 
Building on this foundation, we analyze the key challenges and requirements for interoperability and model composition. 
Through this approach, we aim to provide actionable insights that go beyond theoretical discussions, addressing real-world obstacles in industrial applications.
\section{Methodology}
\label{sec:methodology}
\begin{figure}[h]
    \centering
    \begin{adjustbox}{width=\columnwidth}
        \begin{tikzpicture}
            \filldraw[black] (0,0) circle (0.1cm);
            \draw[ultra thick, ->, decorate, decoration={snake, amplitude=0.4cm, segment length=7cm}] 
                (0,0) -- (12,0);
            
            \draw[thick] (0.6,0.05) -- (0.6,1);
            \node[align=center] at (0.6,1.5) {1. Identify the\\ research objectives};

            \draw[thick] (2.4,0.4) -- (2.4,-1);
            \node[align=center] at (2.4,-1.5) {2. Identify \& characterize\\ target audience};

            \draw[thick] (4.2,-0.1) -- (4.2,1);
            \node[align=center] at (4.2,1.5) {3. Design sampling\\ plan};

            \draw[thick] (6,-0.4) -- (6,-1);
            \node[align=center] at (6,-1.5) {4. Design \& write\\ questionnaire};

            \draw[thick] (7.8,0.1) -- (7.8,1);
            \node[align=center] at (7.8,1.5) {5. Pilot test\\ questionnaire};

            \draw[thick] (9.6,0.35) -- (9.6,-1);
            \node[align=center] at (9.6,-1.5) {6. Distribute\\ questionnaire};

            \draw[thick] (11.4,0) -- (11.4,1);
            \node[align=center] at (11.4,1.5) {7. Analyze results\\ and write report};
            
        \end{tikzpicture}
    \end{adjustbox}
    \caption{Survey research process according to \citet{MarkKasunic.2005}}
    \label{fig:methodology}
\end{figure}
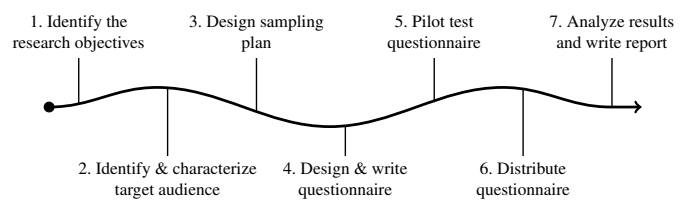

To investigate the \acp{rq} outlined in Sec.~\ref{sec:introduction}, a survey was designed and conducted according to the methodology outlined by \citet{MarkKasunic.2005}, which is displayed in Fig.~\ref{fig:methodology}

As already described in Sec.~\ref{sec:introduction} the target group is composed of experts both from industry and academia.
To ensure the relevance of responses, only professionals with expertise in \acp{dm} are targeted, including scientific staff from partner research institutes, employees of partner companies, and collaborators in research projects, targeting a broad range of sectors and disciplines. 
The questionnaire is primarily disseminated anonymously via email to uphold objectivity and minimize potential bias.
It is structured into three main sections: Persona, Current State, and Desired State. 
This structure aimed to capture both the participants' background information and their insights into the current and future use of \acp{dm} across various lifecycle stages and applications.
\vspace{0.2cm}
\begin{enumerate}[leftmargin=*, nosep]
    \item \textbf{Persona}: The Persona section collected demographic and contextual information about the participants to provide a comprehensive understanding of their professional background. 
    Key aspects included:
    \begin{itemize}[leftmargin=*, nosep]
        \item Field of Work: The participants were asked in which field(s) they are active, allowing multiple selections to reflect interdisciplinary activities.
        \item Industry Sector: The survey inquired about the industry or research area in which the participants operate.
        \item Organizational Context: Information on organizational size (e.g., number of employees) and the participant's role or position was gathered to contextualize their responses.
    \end{itemize}
    \vspace{0.2cm}
    \item \textbf{Current State}: This section explored the participants' current use and management of \acp{dm}, focusing on how these models are employed in their work. 
    Key topics included:
    \begin{itemize}[leftmargin=*, nosep]
        \item Lifecycle Phases: Identification of the lifecycle phases of modeled assets (e.g., design, production, use, operation, maintenance) in which \acp{dm} are utilized.
        \item Use Cases: Participants specified the primary use cases for their \acp{dm}, with options for additional custom use cases not covered in the predefined list.
        \item Types of \acp{dm}: The survey captured the types of models used, differentiating between behavior, structure, and function models \cite{Hildebrandt.2017}, as well as combinations of these.
        \item Model Creation Approaches: Participants were asked whether their \acp{dm} were created using knowledge-driven, data-driven, or hybrid approaches.
        \item Tools and Platforms: Information was gathered on tools and software used for digital modeling, including the use of commercial or open-source platforms and programming libraries (e.g., Python modules).
        \item Standardization: Participants indicated whether they use standardized interfaces (e.g., \ac{fmi}) and data exchange formats, and the level of model interoperability within their workflows.
    \end{itemize}
    \vspace{0.2cm}
    \item \textbf{Desired State}: This section aimed to capture the participants' expectations and desired advancements regarding the future use of \acp{dm}. 
    Key questions addressed:
    \begin{itemize}[leftmargin=*, nosep]
        \item Future Use Cases: Identification of potential new use cases for \acp{dm}, including scenarios that could enhance their work processes.
        \item Missing Models: Participants highlighted gaps in their current \ac{dm} ecosystem, such as unavailable model types or features, and how these limitations could be addressed.
        \item Potential Benefits: Exploration of the benefits participants foresee from the improved integration, synchronization, and modularity of \acp{dm}.
        \item Advanced Technologies: Perspectives on the integration of data-driven methods, artificial intelligence, and enhanced synchronization with real-world assets.
    \end{itemize}
\end{enumerate}
\vspace{0.2cm}
The questionnaire consists of both multiple-choice questions and open-text fields to capture both quantitative and qualitative data. 
Feedback and scoring mechanisms for each question allow participants to elaborate on their choices, providing additional context for the analysis.
It was tested by members of the authors' research institutes before being distributed to about 200 individuals from the authors’ professional network across Germany.

This structured approach enabled the collection of rich data across a wide array of topics, forming the foundation for the validation of the hypotheses and identification of trends in the use of \acp{dm}.

\section{Results}
\label{sec:results}
The following section presents and analyzes the results of the conducted survey on DMs and \acp{dt}. Key findings are visualized through diagrams to support a structured interpretation. A subsequent discussion critically reflects on these insights and delineates the contribution of this study in relation to existing research. 

\subsection{Survey Results} \label{sec:SurveyResults}
\begin{figure} \centering \includegraphics[width=\linewidth]{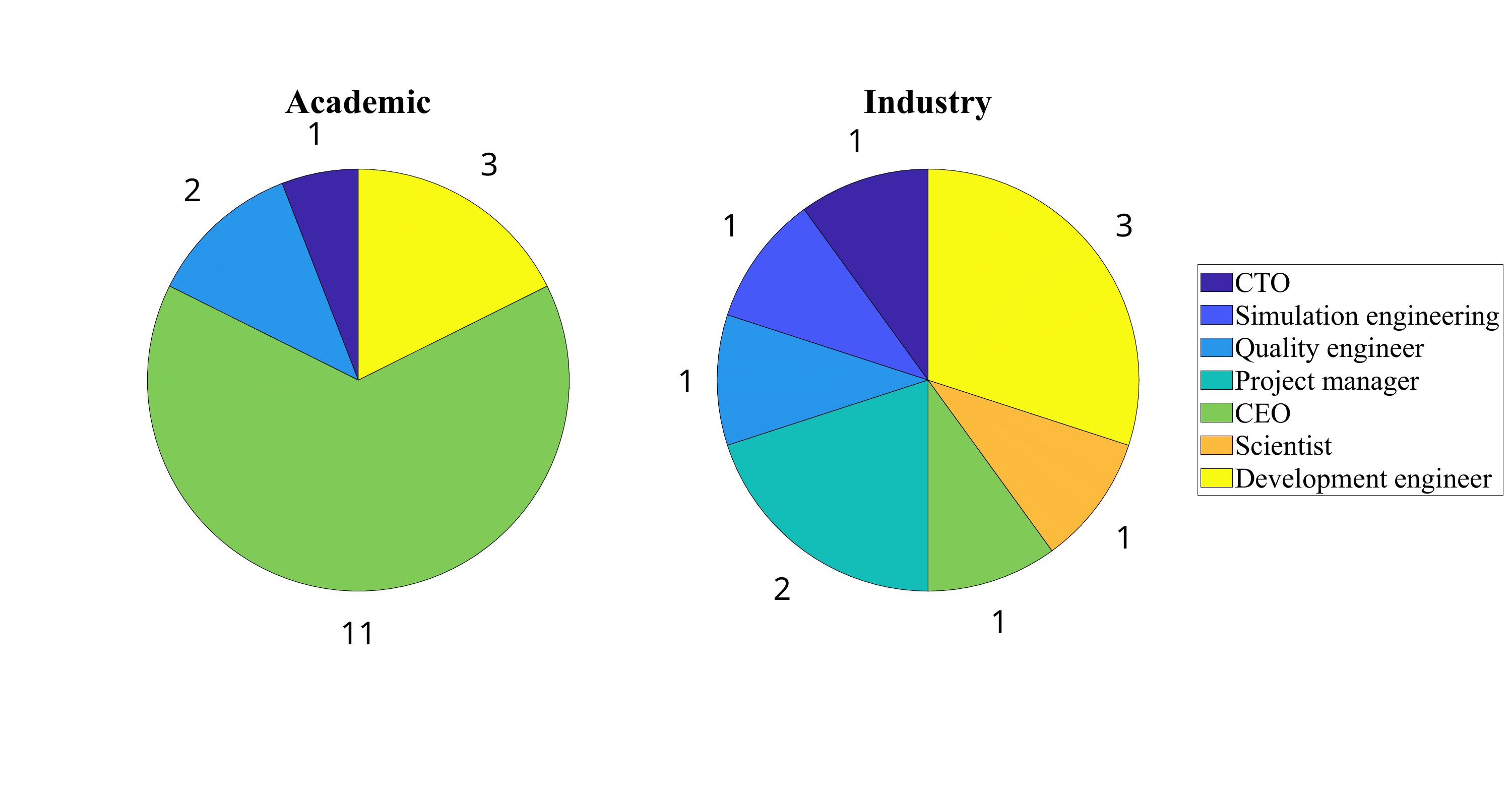} \caption{Distribution of the job positions of the respondents clustered in academic and industry} \label{fig:JobDistribution} \end{figure}

Out of approximately 200 individuals invited to participate, 30 valid responses were obtained. To contextualize the survey results, respondents’ professional backgrounds were categorized into academic and industrial sectors based on their current job roles. 
As illustrated in Fig.~\ref{fig:JobDistribution}, the majority of respondents are from academia (\(n=17\)), while a smaller yet significant proportion come from industry (\(n=10\)). 
Despite this disparity, the survey demonstrates a well-balanced representation of various job roles across both sectors.

\begin{figure} \centering \includegraphics[width=\linewidth]{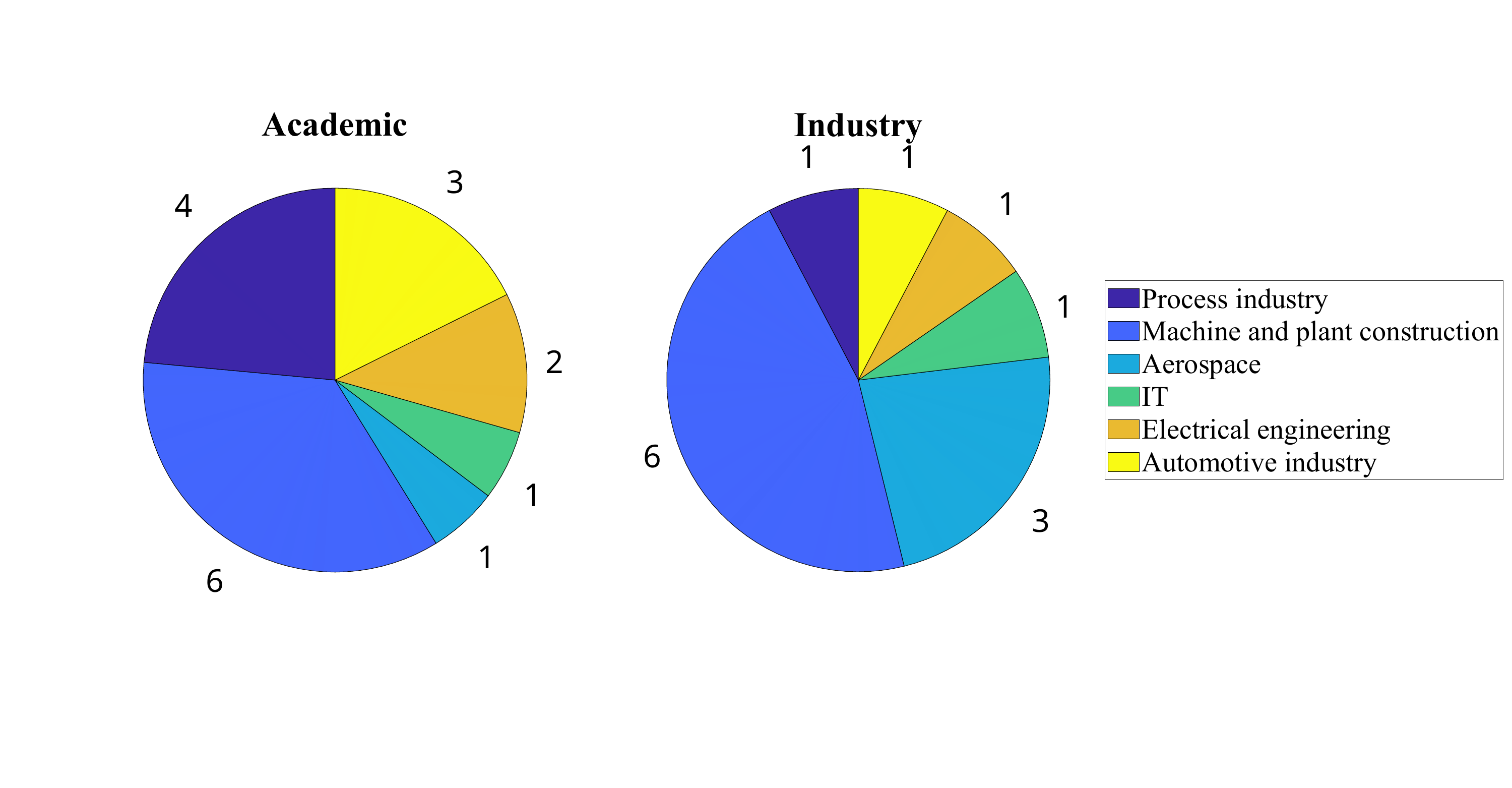} \caption{Distribution of the sectors the respondents are working in clustered in academic and industry} \label{fig:SectorDistribution} \end{figure}

A similar trend can be observed in the sector distribution (Fig.~\ref{fig:SectorDistribution}). 
The majority of respondents work in academic institutions (\(n=17\)), while a smaller proportion are engaged in industry sectors (\(n=13\)). 
Notably, the respondents are involved in a diverse range of domains, indicating a broad applicability of \acp{dm} across both academic and industrial settings.

\begin{figure} \centering \includegraphics[width=\linewidth]{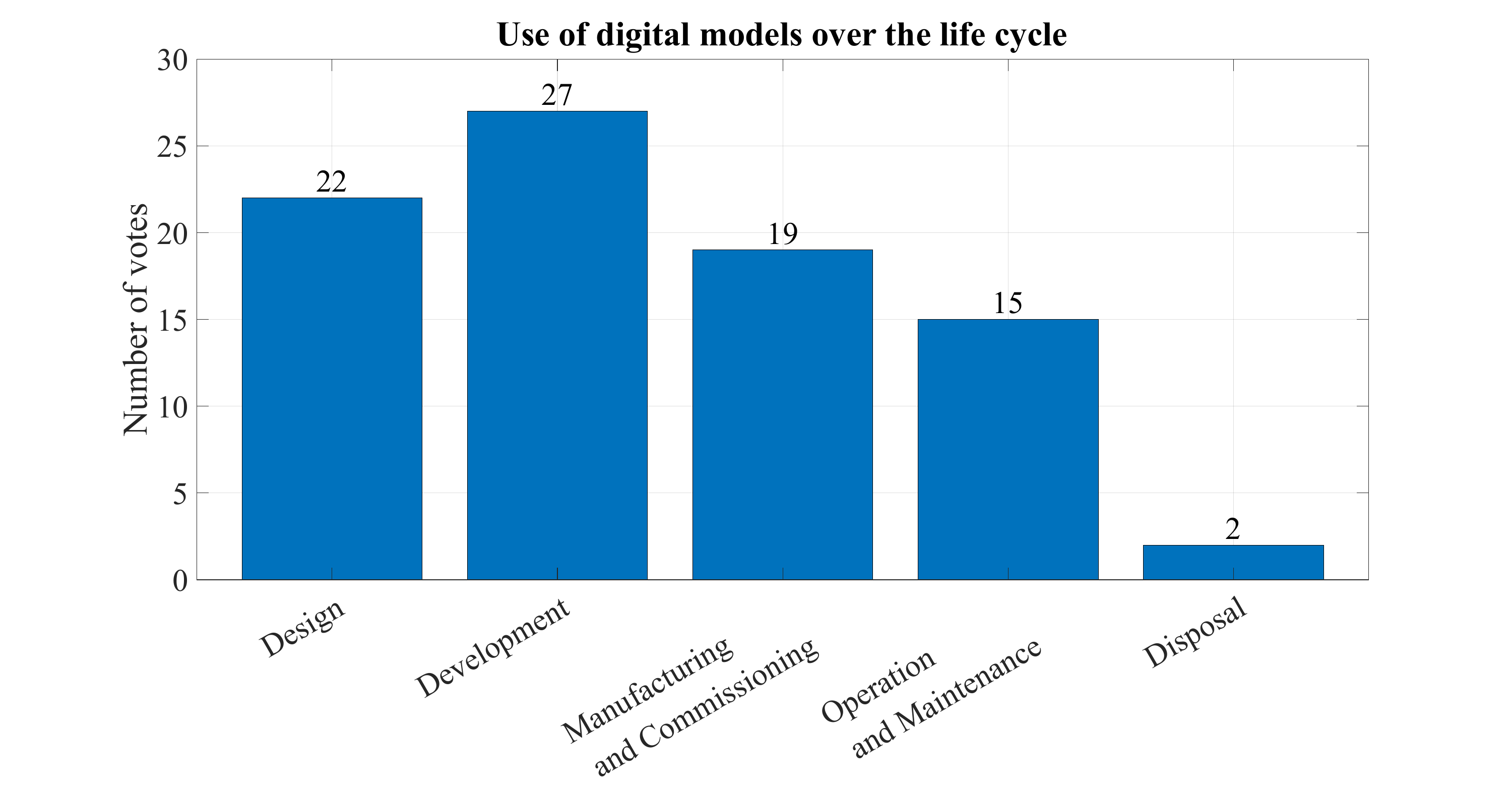} \caption{Usage of \acp{dm} by the respondents in different lifecycle phases} \label{fig:ModelUsageLifecycle} \end{figure}

A key aspect of \ac{dt} applications is the lifecycle phase in which \acp{dm} are used.
Fig.~\ref{fig:ModelUsageLifecycle} shows that \acp{dm} are predominantly utilized in the early lifecycle phases, particularly in design (\(n=22\)) and development (\(n=27\)). 
However, usage significantly declines in later phases, with only two respondents indicating that they employ \acp{dm} during the disposal phase. 
This suggests that models developed in early phases are rarely reused or adapted for later stages, highlighting a potential gap in lifecycle-spanning \ac{dm} utilization.

\begin{figure} \centering \includegraphics[width=\linewidth]{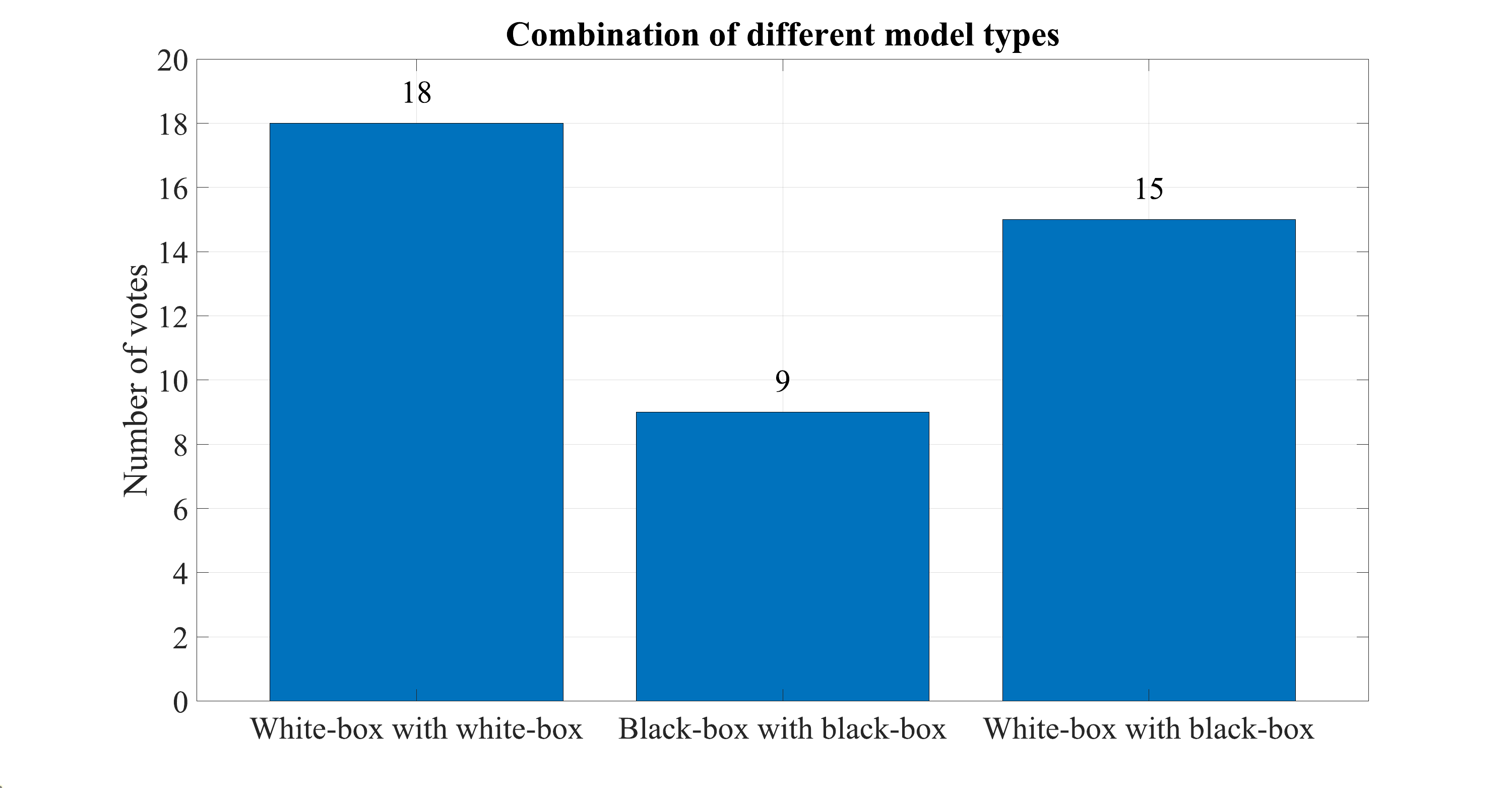} \caption{Combination of different model types} \label{fig:ModelCombination} \end{figure}

Beyond lifecycle considerations, \acp{dm} can be combined in various ways to enhance their functionality. 
Fig.~\ref{fig:ModelCombination} highlights that the most common combination involves white-box models with other white-box models (\(n=18\)), followed by white-box combined with black-box models (\(n=15\)). 
In contrast, the combination of black-box models with each other (\(n=9\)) is less frequent. 
One possible explanation is that white-box models offer greater interpretability, making them easier to analyze and integrate. 
Conversely, black-box models are often more complex, which can introduce challenges in terms of explainability and computational requirements.

\begin{figure} \centering \includegraphics[width=\linewidth]{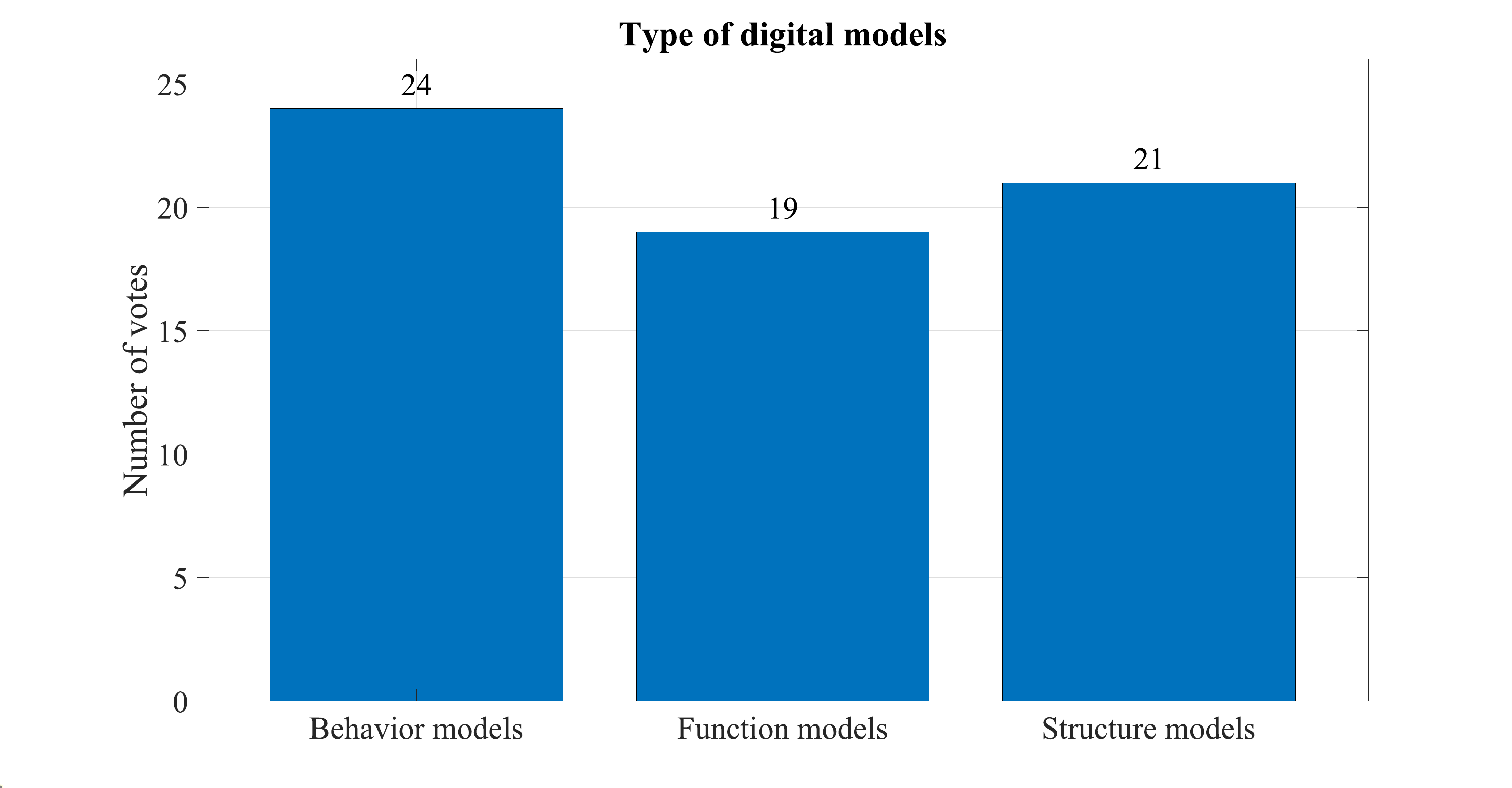} \caption{Types of \acp{dm} the respondents are working with.} \label{fig:ModelTypes} \end{figure}

Examining the types of \acp{dm} employed by respondents (see Fig.~\ref{fig:ModelTypes}), a relatively balanced distribution is evident. 
Behavior models are the most commonly used (\(n=24\), followed closely by structure models (\(n=21\)) and function models (\(n=19\)). 
This suggests that \acp{dm} serve a variety of purposes, from representing system behaviors to structural characteristics and functional aspects.

\begin{figure} \centering \includegraphics[width=\linewidth]{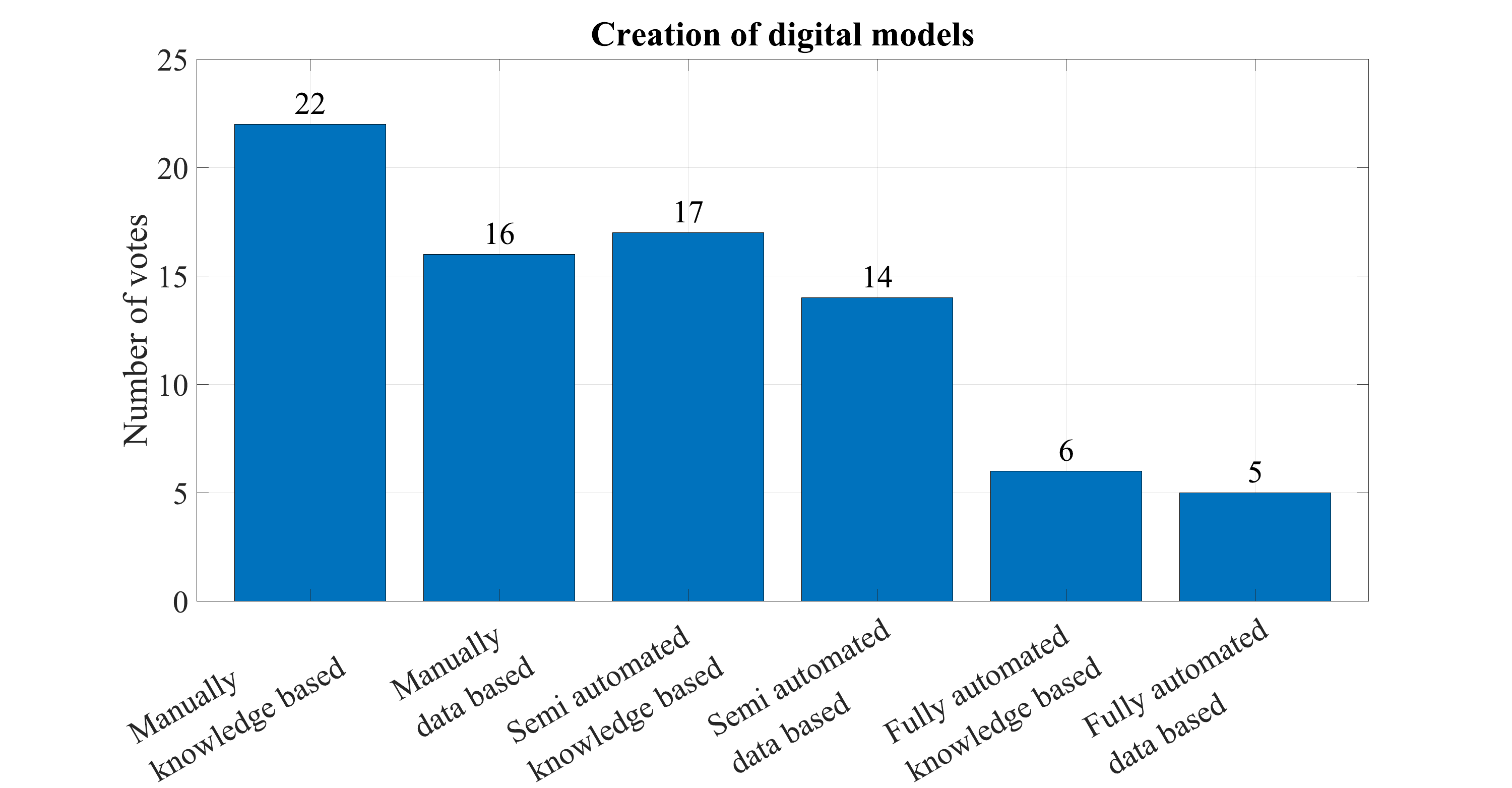} \caption{Approaches used by the respondents in the creation of \acp{dm}.} \label{fig:ModelCreation} \end{figure}

The methodology behind model creation also varies among respondents, as depicted in Fig.~\ref{fig:ModelCreation}.
There is a clear preference for manual model creation compared to semi-automated and automated approaches.
Additionally, within each category, knowledge-based approaches are slightly more favored than data-based ones. 
In the context of the survey, knowledge-based modeling refers to approaches where models are created based on the expertise of the modeler or existing domain knowledge. 
In contrast, data-based modeling describes methods that use large volumes of data to generate models representing specific aspects of a \ac{dt} \cite{VogelHeuser.2021}.
This preference could be attributed to the greater control and flexibility offered by manual creation, allowing experts to directly incorporate domain knowledge and ensure accuracy. 
Furthermore, the collection, processing, and utilization of large datasets for model creation are resource-intensive, which could present an additional obstacle.
In contrast, automated methods may be perceived as less transparent and harder to fine-tune, requiring high-quality input data that may not always be readily available.

\begin{figure} \centering \includegraphics[width=\linewidth]{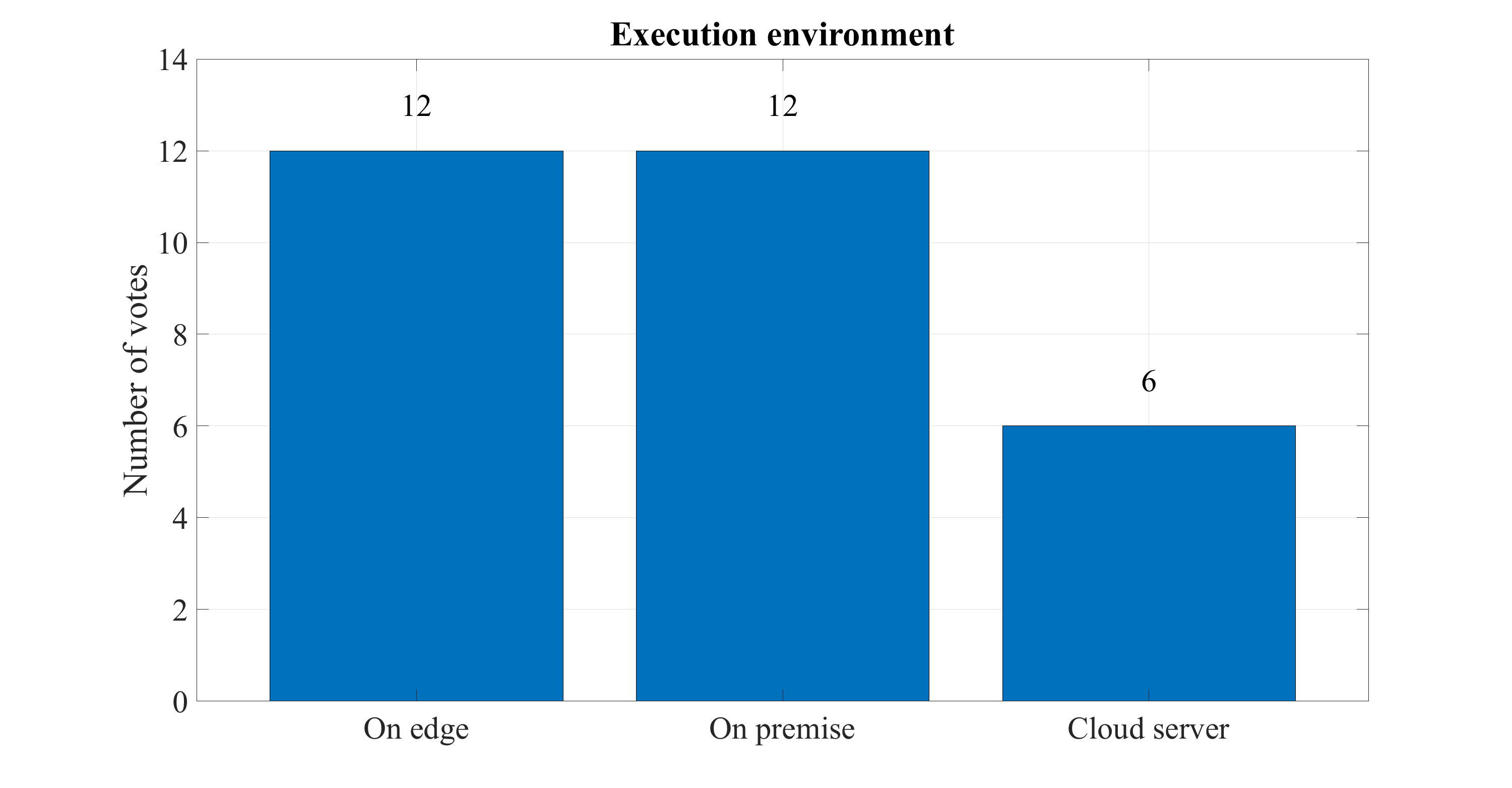} \caption{Execution environments for \acp{dm} used by the respondents} \label{fig:ExecutionEnvironments} \end{figure}

The environments in which \acp{dm} are executed also play a crucial role. 
Fig.~\ref{fig:ExecutionEnvironments} compares different execution environments, showing that edge and on-premise execution are equally preferred (\(n=12\)), whereas cloud-based execution is less common (\(n=6\)). 
One reason for this preference could be concerns regarding intellectual property protection and data security. 
Keeping data locally in edge or on-premise environments enables organizations to maintain control over sensitive information. 
However, as computational demands increase, cloud-based solutions may become more attractive despite existing concerns.

\begin{figure} \centering \includegraphics[width=\linewidth]{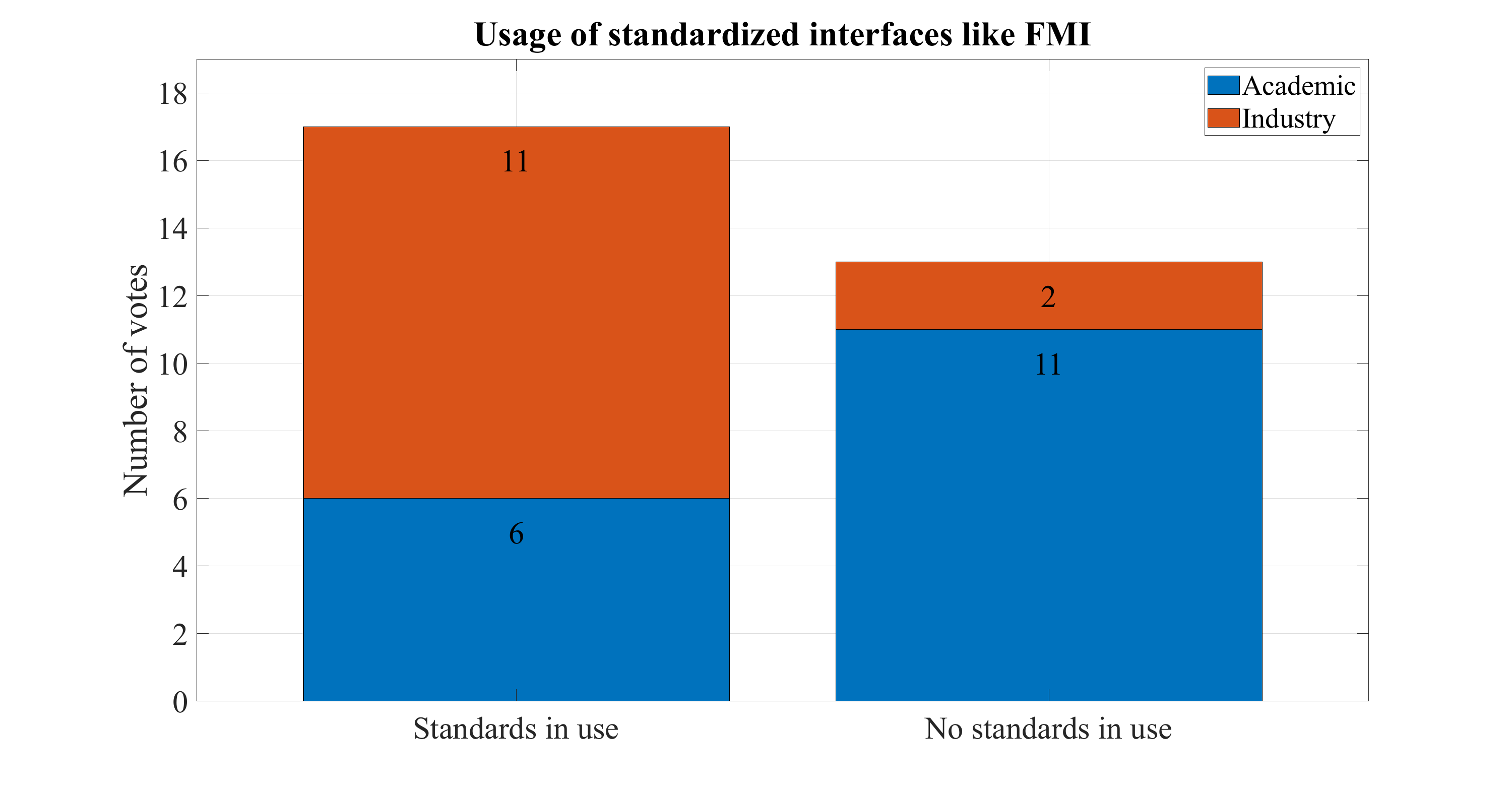} \caption{Usage of standardized interfaces for \acp{dm} like e.g., the \ac{fmi}} \label{fig:Interfaces} \end{figure}

Interoperability is another key aspect of \acp{dm}, often facilitated by standardized interfaces.
Fig.~\ref{fig:Interfaces} illustrates that a majority of respondents (\(n=17\)) use standardized interfaces such as the \ac{fmi}, while a smaller portion (\(n=13\)) do not.
This suggests that standardization is generally well received and used by a significant portion of respondents, potentially improving compatibility and integration across different tools and systems.

\begin{figure} \centering \includegraphics[width=\linewidth]{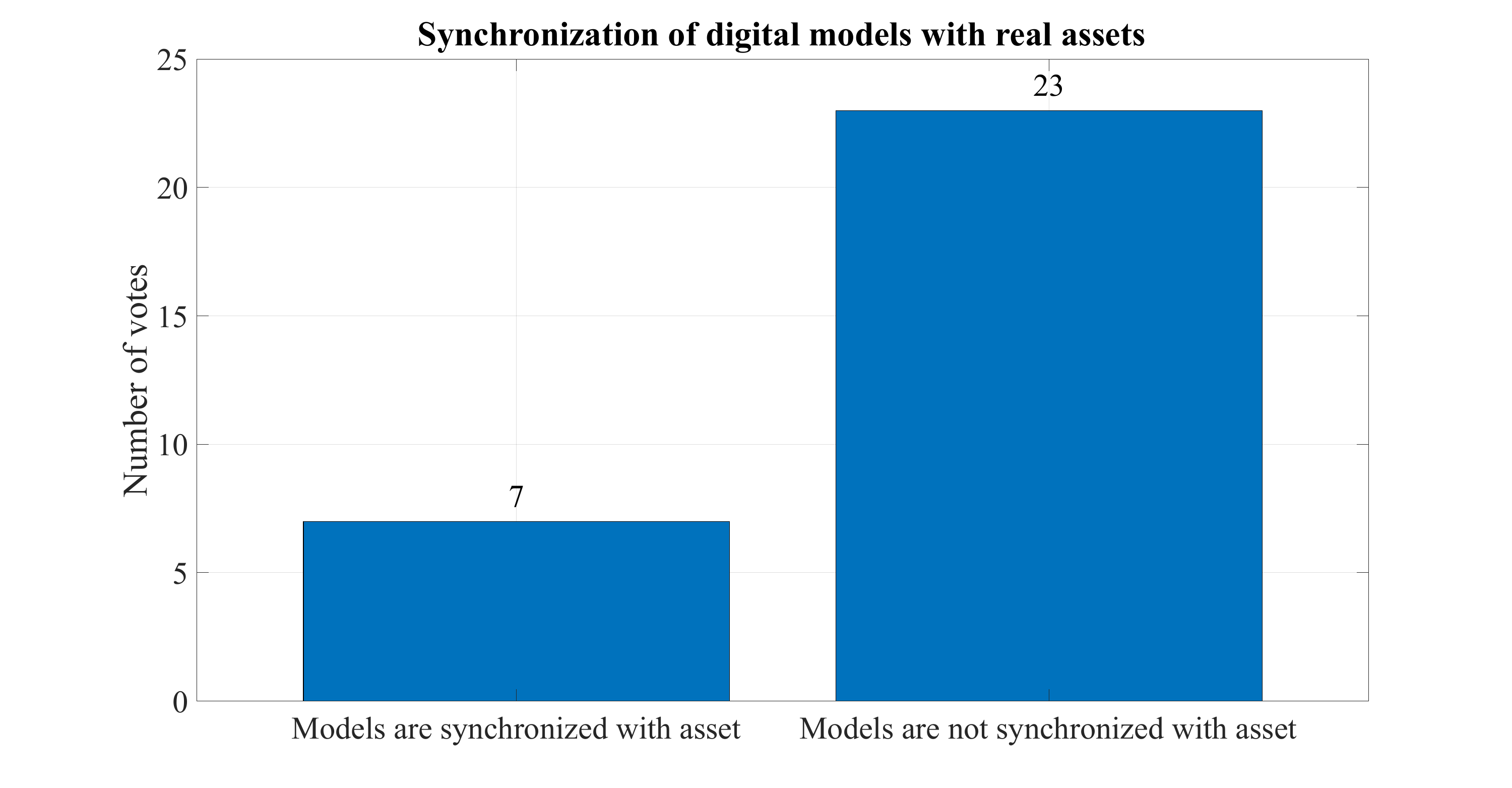} \caption{Distribution in which \acp{dm} are synchronized with physical assets by the respondents} \label{fig:Synchronization} \end{figure}

Finally, the synchronization of \acp{dm} with their corresponding physical assets is an important yet challenging aspect of digital twin implementations. 
As seen in Fig.~\ref{fig:Synchronization}, the majority of respondents (\(n=23\)) do not perform synchronization, whose implementation is seen by various standards as a prerequisite for the implementation of a digital twin \cite{IEC60050-831:2025, IEC30173:2023}, whereas only a few (\(n=7\)) indicate that they do. 
The complexity associated with real-time synchronization—including the need for precise data acquisition, sensor integration, and robust communication protocols—could be a major barrier. 
Additionally, the effort and resources required for continuous synchronization may outweigh the perceived benefits, leading many to rely on periodically updated models instead.

\subsection{Discussion}
Based on the findings presented in Sec. \ref{sec:SurveyResults}, several key challenges can be contextualized and further interpreted.

One possible reason for the discrepancy in the use of \acp{dm} across the lifecycle phases could lie in the current practical application and further development of standardized interfaces.
Heterogeneous data sources and changing framework conditions can make implementation more difficult.
This shows that it is not only the existence of such standards that is decisive, but also their consistent application by the industry.
\acp{oem} in particular play a key role here: they are largely responsible for actively promoting the use of these standards along the value chain and demanding them from suppliers and partners. 
Without this pressure on the entire supply chain, the broad acceptance and implementation of standards is often limited to voluntary initiatives, which can limit their impact. 
In addition, \acp{oem} could help to reduce the challenges of reusing \acp{dm} by increasing their support for standards. 
Currently, adapting and updating existing models often involves considerable manual effort, especially in the later stages of the lifecycle.
This can lead to the use of such models being perceived as uneconomical in practice. 
Stronger institutional anchoring and the clear demand for standards by \acp{oem} could mitigate this by creating incentives for the development of tools and processes that enable more efficient reuse.
In addition to existing standards, it is essential to promote approaches for automatic model generation and adaptation to enable the full potential of \acp{dm} to be exploited throughout the entire lifecycle of an asset. 
The continuous adaptation and extension of models, especially in dynamic and changing operating environments, requires intelligent mechanisms that ensure both rapid model modification and seamless integration of new data and requirements. 
Automated processes for model generation could significantly reduce the burden of manual modeling and updating and thus optimize the reusability of models in different lifecycle phases. 
In particular, the development of adaptive modeling approaches that flexibly adapt to new operating points or disposal conditions could increase the efficiency and cost-effectiveness of model use.
These technologies would reduce complexity and at the same time improve model accuracy and availability. 
Close integration of these approaches with existing standards could not only promote interoperability, but also create the basis for automated, sustainable use of \acp{dm}.

The results of the survey conducted must be critically assessed in view of the limited sample size of only 30 respondents. 
Especially in an area as complex and heterogeneous as the use of \acp{dm} across different lifecycle phases, a broader database should be collected in future work in order to draw a well-founded and far-reaching conclusion. 
However, it should not go unmentioned that the survey covered a broad audience from different industry sectors, which offers a decisive advantage.
By involving experts from different domains, a representative trend could be identified, providing valuable insights into the challenges and potentials of \ac{dm} use. 
The diversity of the participants surveyed ensures that the results are not only viewed from the perspective of a single domain, but also provide a multi-dimensional view of the topic. 
This helps to ensure that, despite the small sample size, a valid trend is presented regarding the acceptance, challenges and wishes in the industry for the use of \acp{dm} across the entire lifecycle. 

\section{Conclusion and Future Work}
\label{sec:conclusion}
The survey conducted in this contribution offers valuable insights into the current application, prevailing challenges, and future expectations regarding the use of \acp{dm} within \acp{dt}. The results reveal a strong presence of \acp{dm} in early lifecycle phases.
More key observations include:
\vspace{0.2cm}
\begin{itemize}[leftmargin=*, nosep] 
    \item \acp{dm} are primarily used during design and development, with limited adoption in maintenance and disposal stages, highlighting the need for improved synchronization mechanisms and adaptive modeling approaches.
    \item A strong preference for manual model creation persists, although hybrid and automated approaches are gaining traction. 
    Increased automation in model adaptation could significantly enhance efficiency and reusability.
    \item Standardization and interoperability remain critical challenges. 
    While some respondents utilize standards like \ac{fmi}, many still rely on proprietary solutions, hindering seamless model integration.
    Strengthening the enforcement of standardized interfaces by \acp{oem} and industry stakeholders could mitigate this issue.
    \item Missing functionalities in \acp{dm}, such as automated model updates, real-time synchronization, and expanded use cases for predictive maintenance and energy efficiency analysis, were identified as key gaps.
\end{itemize}
\vspace{0.2cm}
In the future, the survey is to be expanded in order to obtain more representative data. 
Strengthening standards for seamless model integration and evaluating the use of \ac{ai} methods for their effective and sustainable use in this context will be essential. In addition, improving real-time model synchronization can enhance scalability and efficiency of \ac{dt} implementations while addressing key challenges identified in this study.

\vfill\pagebreak

\section*{Acknowledgements}
This research [projects ProMoDi and LaiLa] is funded by dtec.bw – Digitalization and Technology Research Center of the Bundeswehr. dtec.bw is funded by the European Union – NextGenerationEU.

\bibliography{./bibliography/references.bib}
\bibliographystyle{elsarticle-harv}

\end{document}